\makeatletter
\def\input@path{{styles/}{figs/}}
\makeatother
\documentclass[aip,reprint]{revtex4-2}

\usepackage{xspace}
\usepackage{hyperref}

\usepackage[normalem]{ulem} 
\usepackage{xcolor}

\usepackage{amsfonts}
\usepackage{amsmath}
\usepackage{graphicx}
\usepackage{placeins}
\usepackage{bbm}
\usepackage{xcolor}

\newcommand{\eqn}[1]{Eq.~(\ref{#1})}
\newcommand{\Eqn}[1]{Equation~(\ref{#1})}
\newcommand{\Eqnn}[2]{Equations~(\ref{#1}) and (\ref{#2})}
\newcommand{\eqnn}[2]{Eqs.~(\ref{#1}) and (\ref{#2})}

\begin{document}

\title{Exploiting the path-integral radius of gyration in open quantum dynamics} 

\author{Andrew C.\ Hunt}
\email[]{ach221@cam.ac.uk}
\author{Stuart C.\ Althorpe}
\affiliation{Yusuf Hamied Department of Chemistry, University of Cambridge, Lensfield Road, Cambridge, CB2 1EW, United Kingdom}

\date{\today}

\begin{abstract}
A major challenge in open quantum dynamics is the inclusion of Matsubara-decay terms in the memory kernel, which arise from the quantum-Boltzmann delocalisation of the bath modes. This delocalisation can be quantified by the radius of gyration squared ${\mathcal R}^2(\omega)$  of the imaginary-time Feynman paths of the bath modes as a function of the frequency $\omega$. In a  Hierarchical Equations  of Motion (HEOM) calculation with a Debye--Drude spectral density, ${\mathcal R}^2(\omega)$  is the only quantity that is treated approximately (assuming convergence with respect to hierarchy depth). Here, we show that the well-known Ishizaki--Tanimura correction is equivalent to
 separating smooth from `Brownian' contributions to ${\mathcal R}^2(\omega)$, and that modifying the correction leads to a more efficient HEOM in the case of fast baths. We also develop a simple `A4' adaptation of the `AAA' (Adaptive Antoulas--Anderson) algorithm in order to fit  ${\mathcal R}^2(\omega)$ to a sum over poles, which results in an extremely efficient implementation of the standard HEOM method at low temperatures. 
\end{abstract}


\maketitle 

\section{Introduction}

Open quantum systems are a major research topic in contemporary chemical physics and beyond. The breadth of this field can be gauged by reviews such as refs.~\onlinecite{devegaDynamicsNonMarkovianOpen2017a,tanimuraNumericallyExactApproach2020,delgado-granadosQuantumAlgorithmsApplications2025,nancyrev}, and by recent special issues of this journal.\cite{chinAlgorithmsSoftwareOpen2025,yanfest} The majority of these calculations use generalised forms of the Caldeira--Leggett model,~\cite{caldeiraPathIntegralApproach1983} in which a quantum system is coupled to a bath of harmonic oscillators, with the coupling strength characterised by a spectral density. A vast array of methods has been developed to treat open quantum systems. In chemical physics, widely used numerically exact methods\footnote{The methods cited below are those that have been used in the majority of numerically exact open-quantum calculations reported recently in the chemical physics literature, but are by no means an exhaustive list.} include the Quasiadiabatic Propagator Path Integral (QUAPI),\cite{makriTensorPropagatorIterative1995,makriTensorPropagatorIterative1995a} Multi-Layer Multi-Configuration Time-Dependent Hartree (ML-MCTDH),\cite{wangMultilayerMulticonfigurationTimeDependent2015} Hierarchical Equations of Motion (HEOM)\cite{ishizakiQuantumDynamicsSystem2005a,shiEfficientHierarchicalLiouville2009} and Non-Markovian Quantum State Diffusion (NMQSD)-based methods.\cite{diosiNonMarkovianQuantumState1998,suessHierarchyStochasticPure2014,hartmannExactOpenQuantum2017,lyndCharacterizingRolePeierls2025,stockburgerStochasticLiouvillianAlgorithm1999}  Recent applications have included light-harvesting complexes, \cite{kreisbeckScalableHighPerformanceAlgorithm2014,tongReproducingLowtemperatureExcitation2020,ishizakiTheoreticalExaminationQuantum2009,schroterExcitonVibrationalCoupling2015} molecular spin transport,\cite{faySpinRelaxationRadical2021,fayOriginChiralityInduced2021} and exciton models.\cite{cittyMesoHOPSSizeinvariantScaling2024,varveloFormallyExactSimulations2021,shiEfficientPropagationHierarchical2018}

 A challenge common to all these recent calculations is the inclusion of the extra non-Markovian decay terms in the memory kernel, produced by the quantum statistics of the bath. We will assume below that the bath is bosonic (although analogous treatments have been developed for Fermionic baths\cite{cuiHighlyEfficientAccurate2019,chenUniversalTimedomainProny2022,danEfficientLowtemperatureSimulations2023}), in which case the non-Markovian terms appear as a series, each member of which decays at a successive Matsubara frequency $\omega_n=2n\pi/\beta\hbar$ (where $\beta = 1/k_\text{B}T$). At low temperatures, these terms become especially troublesome, giving a long `Matsubara tail' to the memory kernel. 

In this article, we explore simple ways to treat and interpret the Matsubara-decay terms which exploit the radius of gyration squared ${\mathcal R}^2(\omega)$ of the imaginary-time Feynman paths.
This quantity is a measure of the delocalisation of the bath modes as a function of the frequency $\omega$, and appears directly in the part of the memory kernel responsible for the Matsubara decay. The use of ${\mathcal R}^2(\omega)$ to improve system-bath calculations is therefore not new, and it has been especially useful in the development of the HEOM method. For example, Yan and co-workers have expanded ${\mathcal R}^2(\omega)$ as sums over poles using Pad\'e \cite{huCommunicationPadeSpectrum2010,huPadeSpectrumDecompositions2011} and Fano \cite{cuiHighlyEfficientAccurate2019} approaches, in order to efficiently model the Bose function.  Related work has expanded the memory kernel, in the time \cite{chenUniversalTimedomainProny2022,ikedaGeneralizationHierarchicalEquations2020,tangExtendedHierarchyEquation2015,nakamuraHierarchicalSchrodingerEquations2018,duanZerotemperatureLocalizationSubOhmic2017} and frequency domains.\cite{xuTamingQuantumNoise2022,zhangMinimalPoleRepresentation2025}

However, the interpretation of ${\mathcal R}^2(\omega)$ as the radius of gyration of the Feynman paths is little discussed in the open-quantum literature, and we show below that this simple observation can give rise to interpretational and methodological advantages. This is especially true for HEOM calculations using a Debye--Drude bath, since in this case ${\mathcal R}^2(\omega)$ is the only quantity that is approximated (assuming the user has converged with respect to the hierarchy depth). This article will therefore focus on HEOM calculations with a Debye--Drude bath,\cite{baiHierarchicalEquationsMotion2024a} using the simplest spin-boson model with as a test case. However, our findings are likely to be useful for more complex systems and spectral densities, and perhaps also for open-quantum methods other than HEOM. 

We begin in Sec.~II with a summary of the Caldeira--Leggett formalism, emphasising the role of ${\mathcal R}^2(\omega)$ and explaining its path-integral interpretation. Section~III contains the new material. We show that the path-integral interpretation of ${\mathcal R}^2(\omega)$ gives insight into the well-known Ishizaki--Tanimura low-temperature correction\cite{ishizakiQuantumDynamicsSystem2005} to HEOM, which allows a more accurate modification to be derived. We then show how to fit ${\mathcal R}^2(\omega)$ as a sum over simple poles, which results in a very efficient set of standard HEOM equations. To carry out the fit we adapt the
versatile and powerful `AAA' (Adaptive Antoulas--Anderson) algorithm \cite{nakatsukasaAAAAlgorithmRational2018} which we call the `A4' approach. The AAA algorithm has already been applied within HEOM by Xu et al., who used it to fit the Fourier transform of the memory kernel, obtaining large efficiency savings at cryogenic temperatures for a sub-Ohmic spectral density.\cite{xuTamingQuantumNoise2022} Here, we show that the (A4) fitting of ${\mathcal R}^2(\omega)$ gives comparable efficiency savings for a Debye--Drude spectral density (or other few-pole spectral density) across a broad range of temperatures. In Sec.~IV, we conclude by suggesting some of the ways in which these findings are likely to generalise to non-Debye--Drude spectral densities.

\section{Background theory: the role of ${\boldsymbol{{\mathcal R}^2(\omega)}}$}

In this Section, we summarise some well-known theory for open quantum systems in way that emphasises the role of ${{{\mathcal R}^2(\omega)}}$. To simplify the algebra, we focus mainly on the spin-boson Hamiltonian with a Debye--Drude spectral density, but much of the theory generalises straightforwardly to more complex systems and spectral densities, as indicated below.

The spin-boson Hamiltonian is
\begin{equation}
    \hat H = \hat H_\text{s} + \hat H_\text{b} + \hat \sigma_z \hat F
\end{equation}
where
\begin{gather}
\hat{H}_{\rm s} = \epsilon\hat\sigma_z + \Delta\hat\sigma_x \nonumber\\
    \hat H_\text{b} = \sum_\alpha\frac{\hat p_\alpha^2}{2} + \frac{1}{2}\omega_\alpha^2\hat x_\alpha^2\nonumber\\
    \hat F = \sum_\alpha c_\alpha x_\alpha\label{sob}
\end{gather}
and $\hat\sigma_x$ and $\hat\sigma_z$ are Pauli spin matrices.
The coefficients $c_\alpha$ are obtained from the spectral density $J(\omega)$ using
\begin{equation}
J(\omega) = \frac{\pi}{2}\sum_\alpha \frac{c_\alpha^2}{\omega_\alpha}\left[\delta(\omega-\omega_\alpha)+ \delta(\omega+\omega_\alpha)\right]
\end{equation}

As is usual in the literature, we assume that the density matrix at time $t=0$ is a direct product \cite{breuerOpenQuantumSystems2002} of the form
\begin{equation}
\hat\rho(0)= \hat\rho_{\rm s}(0)\otimes\hat\rho_{\rm b}
\end{equation}
where $\hat\rho_{\rm b}$ is the quantum Boltzmann operator for the bath.
The time evolution of the reduced density
 \begin{equation}\hat \rho_{\rm s}(t) = {\rm tr}_{\rm b}\{e^{-i\hat{H} t/\hbar}\hat\rho(0)\,e^{i\hat{H} t/\hbar}\}
 \end{equation}
 can then be written formally\cite{caldeiraPathIntegralApproach1983} as
  \begin{align}
 \rho_t(s_t,s_t')=\int\! \mathcal{D}{\bf s}\int\! \mathcal{D}{\bf s}' \;\rho_0(s_0,s_0')
F_t[{\bf s},{\bf s}']G_t[{\bf s},{\bf s}']
\end{align}
where $G_t[{\bf s},{\bf s}']$ is the real-time system propagator $e^{-i\hat H_\text{s}t/\hbar}\dots e^{i\hat H_\text{s}t/\hbar}$ expanded over  forward and backward Feynman paths ($s_0\to s_t;s_0'\to s_t'$) (where  $s$ denotes spin up or down in the case of the spin-boson), and $F_t[{\bf s},{\bf s}']$ is the Feynman-Vernon influence functional.\cite{feynmanQuantumMechanicsPath1965,feynmanTheoryGeneralQuantum1963} All we need to know about $F_t[{\bf s},{\bf s}']$ is that it entirely determines the influence of the bath on the system, and that it depends on the bath solely\cite{leggettDynamicsDissipativeTwostate1987} through the memory kernel
\begin{align}
C(t) = {\rm tr}\{\hat\rho_{\rm b}\,\hat F(t)\hat F\}
\end{align}

Most of the challenges in open-quantum calculations can therefore be traced back to properties of the memory kernel, especially differences betwen the quantum kernel $C(t)$ and its classical counterpart $C_\text{class}(t)$. For example,  a Debye--Drude spectral density
\begin{equation}\label{debye}
    J(\omega) = \frac{\eta\gamma\omega}{\omega^2+\gamma^2}
\end{equation}
gives
\begin{align}\label{classy}
C_\text{class}(t) = {\eta\over\beta}e^{-\gamma|t|}
\end{align}
whereas
\begin{equation}
    \label{eq-BCF-debye}
    C(t) = d_0e^{-\gamma |t|} + \sum_{n=1}^\infty d_ne^{-\omega_n |t|}
\end{equation}
where $d_0$=$(\hbar\eta\gamma/2)[\cot (\beta\hbar\gamma/2)-i]$, $d_n$=$-2\eta\gamma\omega_n/[\beta(\gamma^2-\omega_n^2)]$. In addition to a part that decays at the same rate $\gamma$ as the classical kernel, the quantum kernel has a `tail' of extra non-Markovian terms, each decaying at a successive Matsubara frequency $\omega_n=2n\pi/\beta\hbar$. Other spectral densities give rise to similar tails (with the same frequencies $\omega_n$, but different coefficients $d_n$). 

The Matsubara tail is especially problematic for HEOM, since this method  propagates a matrix of system operators\cite{tanimuraNumericallyExactApproach2020}
\begin{equation}\label{ados}
\hat \rho_{\bf m}(t) \equiv \hat \rho_{m_0,m_1,\dots, m_K}(t)
\end{equation}
of which $\hat\rho_{0,0,\dots,0}(t)$ is the reduced density matrix $\hat \rho_\text{s}(t)$, and the other elements are the so-called `auxiliary density operators' (ADOs). Each index $m_n$ runs from zero to $N_n$, where $N_n$ is the depth of the hierarchy associated with the corresponding $n$-th term in \eqn{eq-BCF-debye}, with the sum truncated at some $n=K$. In many calculations (including those we report below in Sec.~III), the $N_n$ are increased until convergence and are independent of $K$. The cost of a HEOM calculation thus grows factorially with $K$, and can become prohibitively expensive at low temperatures. Although this problem is specific to HEOM, other open-quantum methods also become more expensive as the Matsubara tail lengthens at low temperatures.\cite{devegaDynamicsNonMarkovianOpen2017a,nancyrev}

\subsection{Relation of the Matsubara-decay term to ${\boldsymbol{{\mathcal R}^2(\omega)}}$}

The Matsubara terms in \eqn{eq-BCF-debye} are caused by 
the delocalisation of the bath modes, as measured by the radius of gyration squared of the imaginary-time Feynman paths
\begin{align}\label{tiger}
{\mathcal R}^2(\omega) = {\hbar\over 2 \omega}\coth{\beta\hbar\omega\over2} - {1\over \beta\omega^2}
\end{align}
To show this, one writes the bath kernel  $C(t)$ as
\begin{equation}\label{cft}
C(t) = {1\over\pi}\int {\rm d}\omega\, J(\omega) \omega B(\omega,t) 
\end{equation}
where 
\begin{align}\label{bot}
B(\omega,t) =&\langle{\hat x_\omega^2}\rangle\cos\omega t-{i\hbar\over 2\omega}\sin\omega t
\end{align}
is the position autocorrelation function for bath mode $\hat x_\omega$, and
\begin{align}\label{x2}
\langle{\hat x_\omega^2}\rangle = {1\over \beta\omega^2}+{\mathcal R}^2(\omega)
\end{align}
We have written \eqnn{bot}{x2} in this way to emphasise the quantum effects in $B(\omega,t)$, namely the quantum-statistical delocalisation of $\hat x_\omega^2$,  manifested as ${\mathcal R}^2(\omega)$, and the quantum dynamics of the commutator, manifested as the sine term (which is independent of the statistics because $[\hat p_\omega,\hat x_\omega]$ is a constant). Substituting into \eqn{cft}, we obtain
 \begin{align}
    \label{BCF-sp}
    C(t)  
    &= C_\text{class}(t) +\frac{1}{\pi}\int {\rm d}\omega\, J(\omega)\omega{\mathcal R}^2(\omega) \cos\omega t\nonumber\\
    & -\frac{i\hbar}{2\pi}\int {\rm d}\omega\, J(\omega)\sin\omega t
\end{align}
where 
\begin{align}
    \label{BCF-class}
    C_\text{class}(t)  
    &= \frac{1}{\pi\beta}\int {\rm d}\omega\, {J(\omega)\over\omega} \cos\omega t  
\end{align}
is the classical memory kernel. Using the well-known expansion\footnote{Obtained by applying $\coth z=1/z+2z\sum_{n=1}^\infty 1/(z^2+n^2\pi^2)$ to \eqn{tiger}.}
\begin{align}\label{rexpa}
{\mathcal R}^2(\omega) = \lim_{M\to\infty}  {\mathcal R}_M^2(\omega) 
\end{align}
where
\begin{align}\label{rexpa-mats}
 {\mathcal R}_M^2(\omega) ={2\over\beta}\sum_{n=1}^{\overline M}{1\over \omega^2+\omega_n^2}
\end{align}
with $\overline M=(M-1)/2$ ($M$ is odd),\footnote{The truncation is defined to be $\overline M$ rather than $M$ for consistency with Sec.~IIC.} allows one to evaluate the integrals in \eqn{BCF-sp} by 
contour integration, and to show 
 that each of the Matsubara terms in \eqn{eq-BCF-debye} is caused by the corresponding pole in ${\mathcal R}^2(\omega)$.\footnote{The classical term $C_\text{class}(t)$ contributes to $d_0$ because $J(\omega)/\omega$ has a pole at $\omega=-i\gamma$. The two quantum terms also contribute to $d_0$ (for any spectral density $J(\omega)$), since  $J(\omega)$ and $J(\omega)\omega$ have the same poles as $J(\omega)/\omega$, provided $J(\omega)$ is a smooth and odd function of $\omega$. Thus only the poles $\omega=-i\omega_n$ in  ${\mathcal R}^2(\omega)$ contribute to the Matsubara decay coefficients $d_n,n>0$.}

\subsection{Explicit path-integral treatment}
 
For a direct interpretation of ${\mathcal R}^2(\omega)$ and of the individual terms
 in \eqn{rexpa-mats}, we need to represent $\langle \hat x_\omega^2\rangle$ as an imaginary-time path integral. Trotterising the imaginary-time propagator into $P$ equally spaced time slices $\beta_P$ (and dropping the $\omega$ subscript from $\hat x_\omega$), we obtain
 \begin{align} \label{pi1}
 \langle \hat x^2\rangle=\lim_{P\to\infty}{1\over Z_P \cal N}\int {\mathrm d}{\bf x}\, e^{-\beta_P U({\bf x})}\left[{1\over P}\sum_{k=1}^Px_k^2\right]
 \end{align}
 where ${\cal N} = (2\pi\beta_P\hbar^2/m)^{P/2}$, $U({\bf x})$ is the `ring-polymer' potential
  \begin{align} \label{oooo}
  U({\bf x})={1\over 2}\sum_{l=1}^P\left[\omega^2x_l^2 + {(x_{l+1}-x_l)^2\over \beta_P\hbar^2}\right]
   \end{align}
with $x_{P+1}\equiv x_1$, and $Z_P$  is the ring-polymer approximation to the quantum partition function\cite{ceperleyPathIntegralsTheory1995} (obtained by replacing $Z_P$ and the term in square brackets by 1 in \eqn{pi1}). To evaluate the integral, we transform to the normal modes that diagonalise the Hessian of $U({\bf x})$, which (for odd $P$) can be written
\begin{align}\label{MrNormal}
X_0 &= {1\over P}\sum_{l=1}^P x_l\nonumber\\
X_n &= {\sqrt{2}\over P}\sum_{l=1}^P x_l \sin \left({2n\pi l\over P}\right)\nonumber\\
X_{-n} &= {\sqrt{2}\over P}\sum_{l=1}^P x_l \cos \left({2n\pi l\over P}\right)
\end{align}
 where $n=1,\dots,\overline P$, with $\overline P=(P-1)/2$. The `centroid mode' $X_0$ is the centre-of-mass of the polymer, and thus becomes the classical coordinate in the high-temperature limit $\beta\hbar\to0$, in which the polymers collapse to classical points.\cite{althorpePathintegralApproximationsQuantum2021} The $n\ne 0$ modes describe the quantum thermal fluctuations around the centroid at finite temperatures.  
 \Eqn{oooo} then  becomes
 \begin{align} \label{ewe}
  U({\bf X})={P\over 2}\sum_{n=-\overline P}^{\overline P}(\omega^2 + \widetilde \omega_n^2)X_n^2
   \end{align}
where the `ring-polymer' frequencies\cite{craigQuantumStatisticsClassical2004} $\widetilde \omega_n$ are 
\begin{align} \label{rp-frequencies}
 \widetilde \omega_n = {2 \over \beta_P\hbar}\sin\left(n\pi\over P\right)
   \end{align}
and \eqn{pi1} transforms to
  \begin{align} 
  \langle \hat x^2\rangle &= \langle \hat X_0^2\rangle + \sum_{n=1}^{\overline P} \langle \hat X_n^2+ \hat X_{-n}^2\rangle
 \nonumber\\
 &= {1 \over\beta \omega^2} + {\mathcal R}^2_P(\omega)
 \end{align}
 where
 \begin{align}\label{polly}
{\mathcal R}_P^2(\omega) = {2\over\beta}\sum_{n=1}^{\overline P}{1\over \omega^2+\widetilde\omega_n^2}
\end{align}
which is the thermal expectation value of the radius of gyration squared of the $P$-bead ring-polymer. Substituting into \eqn{BCF-sp} for the case of the a Debye--Drude bath (\eqn{debye}) we obtain the $P$-bead approximation to the memory kernel,
  \begin{align}\label{rernel}
C_P(t) = {\widetilde d}_0e^{-\gamma |t|} + \sum_{n=1}^{\overline P} {\widetilde d}_ne^{-\widetilde\omega_n |t|}
  \end{align}
where $\widetilde d_0=\eta/\beta- i\hbar\eta\gamma/2+ (2\gamma^2\eta/\beta)\sum_{n>0}1/(\gamma^2-\widetilde \omega_n^2)$, and $\widetilde d_n=-2\eta\gamma\widetilde{\omega}_n/[\beta(\gamma^2-\widetilde{\omega}_n^2)]$.
 
 Since the path integral is exact in the limit $P\to\infty$, it follows that 
  \begin{align}
    \label{rexpa-rp-infP}
  \lim_{P\to\infty} {\mathcal R}_P^2(\omega) = {\mathcal R}^2(\omega)
  \end{align}
where $ {\mathcal R}^2(\omega)$ is given in \eqnn{tiger}{rexpa}. The expansion in terms of $\widetilde\omega_n$ in \eqn{polly} appears  different to the expansion in terms of  $\omega_n$ in \eqn{rexpa-mats}. However, it can be shown that these two series sum to the same limit, namely ${\mathcal R}^2(\omega)$ of \eqn{tiger}. 

\subsection{Matsubara modes}

The equivalence of \eqnn{rexpa}{rexpa-rp-infP} illustrates a more general property of path integrals. Static averages such as $\langle \hat x^2\rangle$ can be obtained, either by taking the limit $P\to\infty$ directly, or after Fourier-smoothing the ring-polymers,\cite{coalsonConnectionFourierCoefficient1986,ceperleyPathIntegralsTheory1995} such that the beads $x_l,l=1,\dots,P$, become a continuous  function  
 \begin{align}
 x(\tau)=X_0 + \sqrt{2}\sum_{n>0}^{\overline M} X_n\sin{\omega_n \tau} + X_{-n}\cos{\omega_n\tau}
\end{align}
of imaginary time $\tau$.
To carry out the smoothing, one expands $x_l$ in terms of the ring-polymer normal modes
 (of \eqn{MrNormal}), setting the $|n|> \overline M$ modes to zero. One can then take the limit $M\to\infty$ (under the assumption that $M\ll P$). The normal mode frequencies (see \eqn{rp-frequencies}) of the  $M$ Fourier modes thus become
  \begin{align}
 \lim_{P\to\infty}\widetilde \omega_n = {2\pi n\over \beta\hbar} = \omega_n, \quad |n|\le M
 \end{align}
 so the $M$ modes are referred to as the `Matsubara modes'.
 Applying the smoothing to the path-integral in \eqn{pi1} and using \eqn{ewe}, we obtain
\begin{align}  
 {\mathcal R}^2(\omega)=\lim_{M\to\infty} {\mathcal R}_M^2(\omega)
 \end{align} 
 with
 \begin{align}  \label{mamdani}
 {\mathcal R}_M^2(\omega)={\int\!d{\bf X} \left[\prod_{n=-\overline M}^{\overline M} e^{-\beta(\omega^2+\omega_n^2)X_n^2/2}\right]\left[\sum_{m=-{\overline M}}^{\overline M} X_m^2\right]\over\int\!d{\bf X} \left[\prod_{n=-{\overline M}}^{\overline M} e^{-\beta(\omega^2+\omega_n^2)X_n^2/2}\right]}
 \end{align} 
 Tracing back through \eqnn{rexpa}{BCF-sp}, we see that the $n$th Matsubara decay-term in the memory kernel of \eqn{eq-BCF-debye} reflects the contribution made to $ {\mathcal R}^2(\omega)$ by the $\pm n$th Matsubara modes in \eqn{mamdani}. Each Matsubara mode can be thought of as providing its own independent bath,\cite{pradaComparisonMatsubaraDynamics2023d}$^,$\footnote{The smoothed Feynman paths can be shown to follow Newtonian dynamics in the $2M$-dimensional phase space of the $X_n$ and their conjugate momenta: this is an example of `Matsubara dynamics'.}  with  effective spectral density    \begin{align}  
J_n(\omega)= J(\omega){\omega^2\over \omega^2 + \omega_n^2}
 \end{align} 
 The factor of ${\omega^2/(\omega^2 + \omega_n^2)}$, obtained by comparing \eqn{BCF-sp} with \eqn{BCF-class}, reflects the increase in delocalisation of $\hat x_\omega$ with increase in $\omega$.

 \section{Approximating the radius of gyration}
 
  \subsection{Ring-polymer versus Matsubara truncation}
 
 A HEOM calculation requires $ {\mathcal R}^2(\omega)$ to be approximated using a finite number of poles $K$ in order to limit the dimensionality of the ADO matrix of \eqn{ados}. Two such approximations are to use \eqn{polly} with  $\overline P=K$, or \eqn{rexpa} with $\overline M=K$. From Sec.~II we know that both approximations converge to $ {\mathcal R}^2(\omega)$ as $K$ is increased. However, 
it is well known from path-integral simulations\cite{ceperleyPathIntegralsTheory1995,althorpePathintegralApproximationsQuantum2021,ParrinelloRahman1984,ChandlerWolynes1981} that jagged ring-polymers converge much faster than smooth Matsubara paths to exact thermal expectation values. Since this trend is followed by $ {\mathcal R}^2(\omega)$ (Fig.~1) one might expect that the ring polymer expansion leads to a more efficient HEOM than the Matsubara expansion.

\begin{figure}
    \centering
    \includegraphics[scale=1]{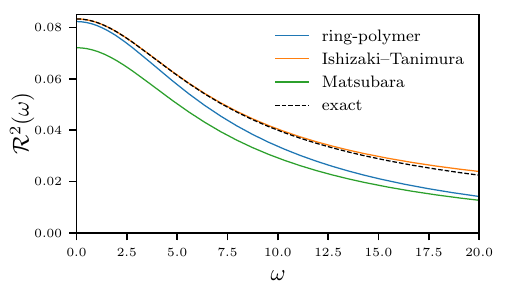}
    \caption{Comparison of the Matsubara and ring-polymer expansions of $ {\mathcal R}^2(\omega)$ (\eqnn{rexpa}{polly}), truncated at $\overline M=\overline P=K=10$ terms, with the Ishizaki--Tanimura (IT)-corrected Matsubara expansion (\eqn{IT-rgapprox}), as a function of the bath frequency $\omega$ (for $\beta=1$).}
    \label{fig1}
\end{figure}

However, a crude truncation of the Matsubara series at $K$ terms is rarely used. Instead, a Markovian Ishizaki--Tanimura (IT) correction term\cite{ishizakiQuantumDynamicsSystem2005} is added, so that $C(t)$ is approximated by
\begin{equation}
    \label{eq-IT-Ct}
   C_{\rm IT}(t)= d_0e^{-\gamma |t|} + \sum_{n=1}^K d_ne^{-\omega_n |t|} + \Gamma_K\delta(t)
\end{equation}
where 
\begin{equation}\label{Kold}
\Gamma_K=-{4\eta\gamma\over\beta}\sum_{n=K\!+\!1}^{\infty} {1\over\gamma^2-\omega_n^2}
\end{equation}
From the perspective of the system, this correction is a dynamical approximation  which assumes that any decay in the memory kernel with a rate faster than $\omega_K$ appears to be instantaneous. From the perspective of the bath, however,  the IT correction is a statistical approximation, equivalent to replacing ${\mathcal R}^2(\omega)$ by 
\begin{equation}
    \label{IT-rgapprox}
    \mathcal{R}^2_{\rm IT}(\omega) =\frac{2}{\beta}\sum_{n=1}^{K}\frac{1}{\omega^2+\omega_n^2} + \frac{2}{\beta}\sum_{n=K\!+\!1}^{\infty}\frac{1}{\omega_n^2-\gamma^2}
\end{equation}
(as is easily verified by substituting $\mathcal{R}^2_{\rm IT}(\omega)$ into \eqn{BCF-sp} and evaluating  the contour integral). The $n>K$ correction term in \eqn{IT-rgapprox} is independent of $\omega$, but this constant term is sufficient to make ${\mathcal R}^2_\text{IT}(\omega)$ a much better approximation than ${\mathcal R}^2_P(\omega)$ to ${\mathcal R}^2(\omega)$---see Fig.~1. 
 As a result, the HEOM obtained from the IT-corrected Matsubara expansion is much more accurate than that obtained from the ring-polymer expansion---see  Fig.~2.
For this reason,  we do not recommend the use of the ring-polymer approximation ${\mathcal R}^2_P(\omega)$, as it results in a less efficient HEOM.
 \begin{figure}
    \centering
    \includegraphics[scale=1]{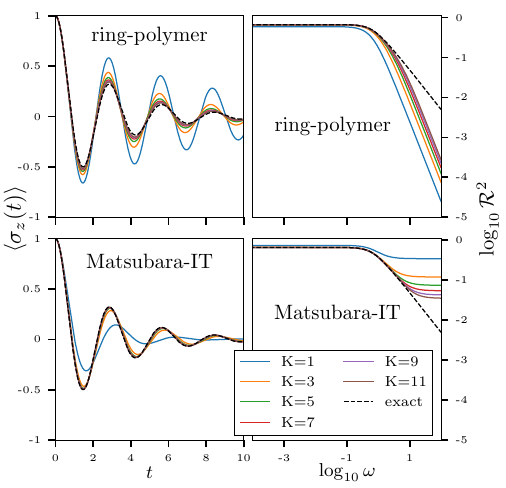}
    \caption{Convergence of the truncated ring-polymer and IT-corrected-Matsubara approximations to ${\mathcal R}^2(\omega)$, and of the resulting HEOM calculations of $\langle \hat \sigma_z(t) \rangle$, as a function of the number of poles $K$. The HEOM calculations were carried out for the spin-boson system of \eqn{sob}, with $\epsilon=0,\Delta=1,\gamma=1,\eta=1$, at $\beta=8$.}
    \label{fig3}
\end{figure}
   
\subsection{Interpretation and modification of the IT correction}

 \Eqnn{mamdani}{IT-rgapprox} show that the IT correction is equivalent to approximating the variances of the $|n|>K$ Matsubara modes by assuming, first, that  $\omega_{K+1}^2\gg\omega^2$, so that the $\omega^2$-dependence can be dropped from the variances, and, second, that $\omega_{K+1}^2\gg\gamma^2$, so that $\gamma^2$ can be subtracted.
  
The first step of this approximation is equivalent to the well-known observation that the high-frequency components of an imaginary-time Feynman path resemble random walks in free space.\cite{ceperleyPathIntegralsTheory1995} These `Brownian' components of the paths resemble a random walk on the lengthscale of the system, as illustrated in Fig.~3, where the $|n|>K$ `Brownian' modes (black line) contribute a noisy `fuzz' to the path, with an amplitude that is small in comparison with the classical variance  $(\beta\omega^2)^{-1}=1$.\begin{figure}
    \centering
    \includegraphics{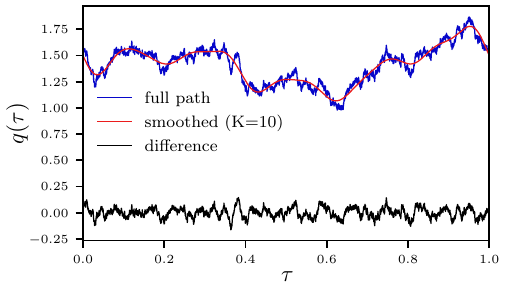}
    \caption{Decomposition of an imaginary-time Feynman path $x(\tau)$ for a bath mode of frequency $\omega=1$,  into smooth $n\le K$ and `Brownian' $n>K$ Matsubara components, for $K=10$ and  $\beta=1$. Note the low amplitude of the Brownian path in comparison with the classical variance (=1).}
    \label{fig:ITfig}
\end{figure}

The second step, in which $\gamma^2$ is subtracted from the variance, is therefore unnecessary. To exploit the Brownian nature of the $|n|>K$ components one only needs to drop the $\omega^2$ contribution to the variance, approximating $\mathcal{R}^2(\omega)$ by
\begin{equation}
    \label{PA-rgapprox}
  \mathcal{R}^2_{\rm mIT}(\omega) =\frac{2}{\beta}\sum_{n=1}^{K}\frac{1}{\omega^2+\omega_n^2} + \frac{2}{\beta}\sum_{n=K\!+\!1}^{\infty}\frac{1}{\omega_n^2}
\end{equation}
which results in a modified IT (mIT) correction, with $\Gamma_K$ in  \eqn{eq-IT-Ct}  replaced by
\begin{equation}\label{Kmart}
   \Gamma_K'=-{4\eta\gamma\over\beta}\sum_{n=K\!+\!1}^{\infty} {1\over\omega_n^2}
    \end{equation}
Evaluation of the leading-order error term 
 \begin{equation}
    \mathcal{R}^2(\omega)\approx\mathcal{R}^2_{\rm IT}(\omega) +  \mathcal{O}\!\left(\frac{\omega^2+\gamma^2}{\omega^2+\omega_{K+1}^2}\right)
 \end{equation}
 confirms that the modified IT approximation is more accurate than its original form. Furthermore,  the modification has removed the unphysical dependence of the radius of gyration (a system-independent property) on the bath parameter $\gamma$ (compare \eqnn{IT-rgapprox}{PA-rgapprox}).
 
Whether this modification gives a significant reduction in the number of ADOs is dependent on the speed of the bath. If the bath is slow enough that $\omega_\text{max}\gtrsim \gamma$ (where $\omega_\text{max}$ is the largest value of $\omega$ to contribute significantly to the integral in \eqn{BCF-sp}), then the condition 
$\omega_{K+1}^2\gg\omega^2$ implies $\omega_{K+1}^2\gg\gamma^2$, so one expects no noticeable improvement on replacing
$\Gamma_K$ by $\Gamma_K'$. The spin-boson results of Fig.~2 are an example of such a bath (for which the mIT results agree with the IT results to within graphical accuracy and are therefore not shown).

 However, if the bath is fast enough that $\gamma^2\gg \omega_\text{max}^2$
 (i.e.~if decay-rates significantly smaller than $\gamma$ appear instantaneous to the system)  then replacing
$\Gamma_K$ by $\Gamma_K'$ should improve the convergence of the HEOM,
since $\omega_K$ can be set to a value determined by $\omega_\text{max}$ rather than $\gamma$.
Figure~4 gives an example of such a bath, where the switch from IT to mIT results in quicker and cleaner convergence of ${\mathcal R}^2(\omega)$ and of the resulting HEOM calculations. The value of $\gamma$ in Fig.~4 was chosen to avoid a fortuitous `resonance' of $\gamma$ with one of the Matsubara frequencies $\omega_n$. In the case that $\gamma\approx\omega_n$, the modification gives a drastic improvement in efficiency---see Appendix A. For these reasons, we
recommend that the modified form of the Ishizaki--Tanimura truncation correction (\eqn{Kmart}) be used in preference to its original form (\eqn{Kold}). In addition, the system-independence of $\mathcal{R}^2_{\rm mIT}(\omega)$ ensures that it can be substituted into \eqn{BCF-class} to derive analogous truncation corrections for any form of spectral density $J(\omega)$.

\begin{figure}
    \centering
    \includegraphics[scale=1]{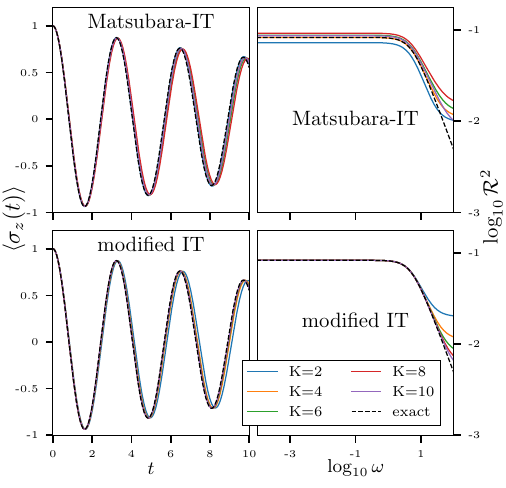}
    \caption{Demonstration that the modified version of the IT correction term (\eqn{Kmart}) gives cleaner convergence for a fast bath than the original form (\eqn{Kold}). All parameters are the same as in Fig.~2 except that $\gamma=61.3$ and $\beta=1$.}
\end{figure}
 
 \subsection{Fitting ${\boldsymbol{{\mathcal R}^2(\omega)}}$ using an `A4' adaption of the AAA algorithm}
 Equations~(\ref{rexpa-mats}), (\ref{polly}), (\ref{IT-rgapprox}) and (\ref{PA-rgapprox}) are examples of sum-over-poles approximations to $\mathcal{R}^2(\omega)$, of the form
  \begin{equation}
    \label{A4rg} \mathcal{R}^2(\omega) \approx \sum_{n=1}^K \frac{k_n}{\omega^2+\eta_n^2} + k_0
 \end{equation}
 in which the poles are $\pm i \eta_n$, and the coefficients $k_n$ are $\omega$-independent constants. In the examples seen so far, $\eta_n$ have been set as either $\widetilde\omega_n$ or $\omega_n$, but there are a variety of other methods in the HEOM literature which approximate $\mathcal{R}^2(\omega)$ using a different choice of $\eta_n$ in order to obtain better accuracy. All such methods can be regarded equivalently as fits to the Bose function $n_\beta(\omega)$, to which $\mathcal{R}^2(\omega)$ is related by 
  \begin{equation}
    n_\beta(\omega) = \frac{1}{\beta\hbar\omega}+\frac{\omega}{\hbar}\mathcal{R}^2(\omega) -\frac{1}{2}
 \end{equation}
 For example, the powerful [N/N] Pad\'e technique\cite{huCommunicationPadeSpectrum2010,huPadeSpectrumDecompositions2011} of Yan and co-workers,\footnote{There are other varieties of the Pad\'e method which include the [N-1/N] version (in which $k_0=0$) and the [N+1/N] version (in which an $\omega^2$ term is added to \eqn{A4rg}). So far as we can determine, [N/N] Pad\'e gives the best results for a Debye--Drude bath.} determines $\eta_n$ and $k_n$ by effectively Taylor expanding $\mathcal{R}^2(\omega)$ about $\omega=0$. Figure~5 illustrates how well the use of the Pad\'e-determined $\eta_n$ and $k_n$ improve the accuracy of $\mathcal{R}^2(\omega)$ and the HEOM results compared with the IT-truncated Matsubara expansion. 
  
 \begin{figure}
    \centering
    \includegraphics[scale=1]{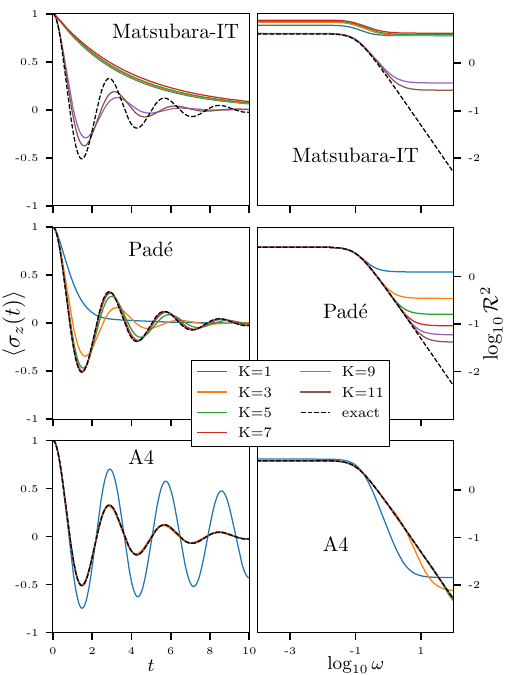}
    \caption{Convergence of the IT-corrected Matsubara, [N/N] Pad\'e, and A4 approximations to $\mathcal{R}^2(\omega)$ and of the resulting HEOM calculations, with respect to the number of terms $K$ in the sum-over-poles expansion. The system and bath parameters are the same as in Fig.~2 and $\beta=50$.
}
\end{figure}

  \begin{figure}
    \centering
    \includegraphics[scale=1]{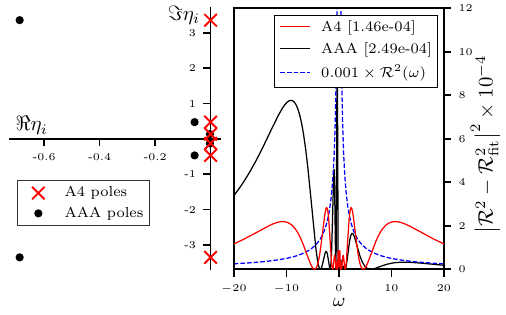}
    \caption{Same as Fig.~5, but with $\beta=500$.}
\end{figure}

The appearance of $\mathcal{R}^2(\omega)$ in \eqn{BCF-sp} implies that even better results than those of the [N/N] Pad\'e method would be obtained if one could fit $\mathcal{R}^2(\omega)$ to an expansion of the form of \eqn{A4rg}, over the full range of $\omega=[-\omega_\text{max},\omega_\text{max}]$ that contributes to the memory kernel.
Carrying out such a non-linear fit can be challenging. An interesting recursive algorithm is proposed in ref.~\onlinecite{cuiHighlyEfficientAccurate2019} which demonstrates that fitting over the full range of $\omega$ does indeed give marked improvement over the Pad\'e approach. However, this algorithm requires fitting parameters to be specified by the user, and can result in higher-order poles in the expansion which lead to a more complicated HEOM. Here, we suggest a simple, direct, fit which is based on the AAA algorithm, which we will refer to as the `A4' approach.

The AAA algorithm fits any rational function to an expansion of the form 
\begin{equation}\label{aaa}
f(\omega) \approx z_0 + \sum_{j=1}^{2K} \frac{z_j}{\omega-\zeta_j}
\end{equation}
where the poles $\zeta_j$ and the residues $z_j$ are complex. The only input required is a grid of values of $f(\omega)$ evaluated over the `support' (i.e. the domain in the complex plane over which one wishes to fit $f(\omega)$), and the desired tolerance $\tau_\text{tol}$; the number of poles $2K$ in the expansion is then determined automatically by the algorithm.
The algorithm is stable, fast and easy to implement (using a MATLAB routine distributed by the authors of ref.~\onlinecite{nakatsukasaAAAAlgorithmRational2018} or its Python implementation in SciPy\cite{virtanenSciPy10Fundamental2020}). It has been applied across a broad range of disciplines, including HEOM: an impressive calculation by Xu et al.\ used the AAA algorithm to fit $S_\beta(\omega)$ (i.e.\ the Fourier transform of $C(t)$) to the form of \eqn{aaa}, which then allowed HEOM calculations to be done at unprecedentedly low temperatures for a subohmic $J(\omega)$ containing a large number of poles.\cite{xuTamingQuantumNoise2022} This efficiency was achieved by merging the poles in $J(\omega)$ with the poles in $\mathcal{R}^2(\omega)$. However, for a Debye--Drude spectral density (for which $J(\omega)$ has only one pole) no reduction in the number of poles would be obtained by fitting $S_\beta(\omega)$ instead of $\mathcal{R}^2(\omega)$. In fact, the resulting HEOM would be a lot more expensive, since $S_\beta(\omega)$ has complex poles, which doubles the dimensionality of the resulting ADO matrix (with respect to the matrix obtained from the same number of real poles).
 
\begin{figure}
    \centering
    \includegraphics[scale=1]{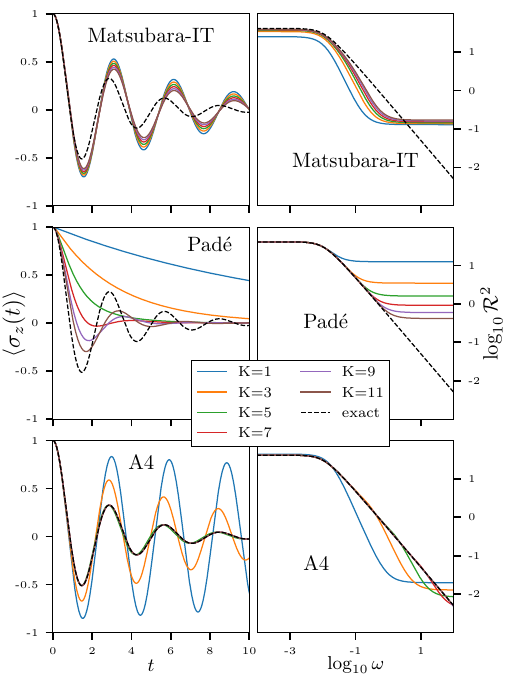}
    \caption{Comparison of A4 and raw AAA fits of $\mathcal{R}^2(\omega)$ for $\beta=50$, $K=3$, showing the pole locations in the complex plane (left panel) and the fitting errors (right panel). The mean-square errors of the fits are given in square brackets.}
    \label{fig:A4example}
\end{figure}

We therefore need to adapt the AAA algorithm so that it fits $\mathcal{R}^2(\omega)$ to a set of purely imaginary poles---i.e.\ to the form of \eqn{A4rg}. It turns out that an out-of-the-box application of AAA almost does this job, but not quite. On applying AAA with a value of $\tau_\text{tol}$ sufficient to fit $\mathcal{R}^2(\omega)$ to graphical accuracy over a dense, equally spaced, grid of support points, distributed in the interval $\omega\in[-\omega_\text{lim},\omega_\text{lim}]$, we find that the poles $\zeta_j$ are complex, but the imaginary parts are typically much larger than the real parts. The poles are not located in groups of four ($\pm a \pm i b$), as one might expect given that $\mathcal{R}^2(\omega)$ is symmetric under $\omega\to-\omega$; instead they lie on the positive side of the real axis (in pairs $a \pm i b$), suggesting that the real parts of the poles are probably spurious, at least for small $K$ (although for some high-$K$ cases that we tested, the real parts of the poles combined in such a way as to improve the accuracy of the fit slightly over the range of the support). It therefore makes sense to neglect the real parts of the poles, then to refit, so the A4 approach is as follows: 
 \begin{enumerate}
\item Carry out an AAA fit of $\mathcal{R}^2(\omega)$ using a dense, equally spaced, grid of support points over $\omega=[-\omega_\text{lim},\omega_\text{lim}]$, with $\tau_\text{tol}$ set to the smallest value that yields $2K$ poles in the expansion of \eqn{aaa}.
\item Discard the residues $z_j$ and the real components of the poles $\zeta_j$, then use the purely imaginary components as the poles $\pm \eta_n$ in \eqn{A4rg}, and determine the coefficients $k_n$ using a linear least-squares fit.
\end{enumerate}
A comparison of an A4 fit (at $\beta= 50$ with $K = 3$) with the corresponding raw AAA fit  is shown in Fig.~7. By discarding the real parts of the AAA complex poles, the A4 procedure has restored the $\omega\to-\omega$ symmetry to the fit, and has also reduced the mean square error. Similar results were obtained at all other temperatures tested (from $\beta=1\to 500$).
  
We have tested the A4 approach in HEOM calculations for the spin-boson system of \eqn{sob}, over a broad range of temperatures ($\beta=1\to 500$). Figures 5 and 6 show the results obtained at $\beta=50$ and $\beta=500$. We used $10^5$ support points with $\omega_\text{lim}=200$ (which was chosen to be sufficiently large that $J(\omega_\text{lim})$ is equal to 1\% of its maximum value). An easy-to-use Python implementation of the A4 approach is included in the supplementary material, and is also publicly available on GitHub\cite{hunt2026A4}. 

The A4 approach strongly and consistently  outperforms the Pad\'e method, and this advantage accelerates as $\beta$ is increased. By $\beta=50$ (Fig.~5),  the A4 approach converges much more rapidly with $K$ than the Pad\'e method. By $\beta=500$ (Fig.~6), 
HEOM calculations can be converged easily using A4 (on a cheap laptop), whereas comparable calculations using the Pad\'e approach would be orders of magnitude more costly.

 \ 
 
 \section{Conclusions}
 
The radius of gyration $\mathcal{R}^2(\omega)$ is solely responsible for the Matsubara terms in the memory kernel, and  is the only quantity that is approximated in a HEOM calculation with a Debye--Drude bath (assuming convergence with respect to hierarchy depth). Not surprisingly, therefore, the success or failure of such a calculation depends entirely on the approximation made to $\mathcal{R}^2(\omega)$.

We have shown here that the Ishizaki--Tanimura correction approximates $\mathcal{R}^2(\omega)$ in a very interesting way, by separating out the noisy `Brownian' components of imaginary-time Feynman paths of the bath. When one realises this, one can modify the correction to make it more efficient in the case of fast baths.
 
At low temperatures, it is far more efficient to approximate $\mathcal{R}^2(\omega)$ by fitting it to a sum over poles, as is well known from the success of the Pad\'e method.\cite{huCommunicationPadeSpectrum2010,huPadeSpectrumDecompositions2011} However, the latter fits $\mathcal{R}^2(\omega)$ about $\omega = 0$, whereas much better results are obtained by fitting over the full range of values of $\omega$ that contribute to the memory kernel. Previous work\cite{cuiHighlyEfficientAccurate2019} reported such a fit using an iterative approach to a sum over higher-order poles (which results in a generalised HEOM). Here, we show that a direct fit to a sum over simple poles can be done using an `A4' modification of the AAA algorithm, which results in the standard HEOM and is orders of magnitude more efficient than the Pad\'e approach at low temperatures. The A4 fit is easy to implement using the Python script given in the supplementary material. 

Further work will be required to assess how well the A4 approach works for non-Debye baths. At near-zero temperatures with sub-Ohmic spectral densities, it is likely that approaches which fit the entire bath kernel\cite{chenUniversalTimedomainProny2022,tangExtendedHierarchyEquation2015,ikedaGeneralizationHierarchicalEquations2020,xuTamingQuantumNoise2022} will overtake the A4 approach in efficiency. But at non-cryogenic temperatures with spectral densities that can be represented in terms of relatively few poles (e.g., using the Meier--Tannor approach \cite{meierNonMarkovianEvolutionDensity1999}), the A4 approach is expected to be very efficient. 

The A4 approach exploits a fortunate and unexpected feature of the AAA algorithm, namely that it returns poles with small real parts when used to fit $\mathcal{R}^2(\omega)$.\footnote{This is not simply a consequence of the $\omega\to-\omega$ symmetry of $\mathcal{R}^2(\omega)$: the use of AAA to fit other symmetric functions often yields complex poles with large real components.}  This feature is also found when the AAA algorithm is applied to the Fermionic analogue of $\mathcal{R}^2(\omega)$, indicating that the A4 approach should also lead to very efficient HEOM calculations for treating Fermionic  baths. However, it is possible that some of the other non-linear fitting techniques available (e.g. those discussed in ref.~\onlinecite{takahashiHighAccuracyExponential2024}) also have this feature. We plan to investigate these possibilities in future work.

\section*{{Supplementary Material}}

See the supplementary material for an easy-to-use Python implementation of the A4 approach for the decomposition of both Bose and Fermi functions. The code is also available on GitHub. \cite{hunt2026A4}

\section*{Acknowledgements}

ACH acknowledges support from the Engineering and Physical Sciences Research Council (EPSRC) and the Yusuf Hamied Fund through a Doctoral Training Account (DTA) PhD studentship, and from Gonville and Caius College, University of Cambridge.
\section*{Author Declarations}
\subsection*{Conflict of Interest}
The authors have no conflicts of interest to disclose.
\subsection*{Author Contributions}
\noindent
\textbf{A.~C.~Hunt}: Conceptualization (equal); Formal analysis (lead); Investigation (lead); Software (lead); Writing – original draft (lead); Writing – review \& editing (equal). \textbf{S.~C.~Althorpe}: Conceptualization (equal); Formal analysis (supporting); Investigation (supporting); Writing – review \& editing (equal).

\section*{Data Availability}
Data presented in the paper are available from the corresponding author upon reasonable request. The code used to perform the simulations is publicly available on GitHub.\cite{hunt2025Feom}

\appendix

\section{Singularities in the Ishizaki-Tanimura truncation correction}

The IT correction term $\Gamma_K$ of \eqn{Kold} is singular if $\gamma=\omega_n$ for one of the terms in the sum. As result, the modification to the IT correction proposed in Sec.~IIB, whereby  $\Gamma_K$ is replaced by $\Gamma_K'$, leads to a drastic improvement in the approximation to $\mathcal{R}^2(\omega)$ and hence in the efficiency of the resulting HEOM if the bath is fast (i.e.\ $\gamma\gg\omega_\text{max}$) and if $\gamma$ is close to one of the Matsubara frequencies for which $\omega_n>\omega_\text{max}$.

This behaviour is illustrated in Fig.~8, which uses the same parameters as the calculations of Fig.~4, except that $\gamma$ has been tweaked to bring it close to resonance with $\omega_{10}$. For $K<10$,  $\mathcal{R}_\text{IT}^2(\omega)$ is spuriously large and the resulting HEOM calculation is thus very far from convergence; only when $K\ge 10$ (such that $\Gamma_K$ no longer includes the $n=10$ term) are the IT results reasonable. The mIT calculations, by contrast, do not suffer from this problem, and yield a good approximation to the HEOM results with just $K=2$, and convergence to within graphical accuracy with $K=4$.\footnote{Note that the same $\gamma=\omega_n$ singularity appears in the $d_n$ coefficient of \eqn{eq-BCF-debye} as in $\Gamma_K$, but is cancelled out by a term in the expansion of the $\cot(\beta\hbar\gamma/2)$ term in $d_0$, thus ensuring that the resulting HEOM calculations are well converged. This is another example of the general observation that a good approximation to $\mathcal{R}^2(\omega)$ results in a well-converged HEOM.}

\begin{figure}
    \centering
    \includegraphics[scale=1]{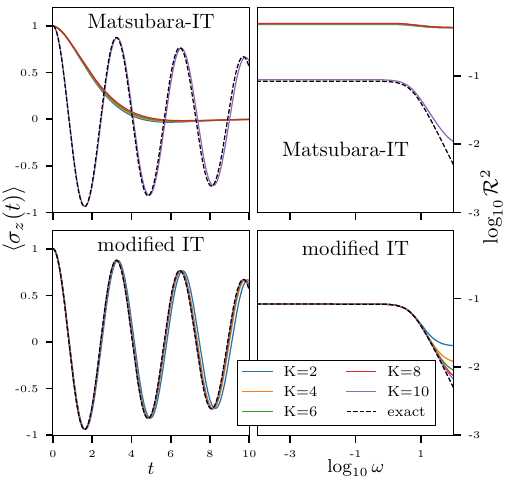}
    \caption{Same as Fig.~4, except that $\gamma$ has been tuned so that $\gamma=62.8\approx\omega_{10}$.}
\end{figure}


%
%

%

\FloatBarrier
\section*{References}
\bibliography{refs}

\end{document}